\documentclass{article}

\usepackage{amsmath,bm,graphicx}
\usepackage{cite}

\newcommand{\rd}{\mathrm{d}}
\newcommand{\ri}{\mathrm{i}}
\begin{document}
\begin{titlepage}
\begin{center}
{\large \textbf{Autonomous models on a Cayley tree}}

\vspace{2\baselineskip}
{\sffamily Mohammad~Khorrami~\footnote{e-mail: mamwad@mailaps.org},
Amir~Aghamohammadi~\footnote{e-mail: mohamadi@alzahra.ac.ir}.}

\vspace{\baselineskip}
Department of Physics, Alzahra University, Tehran 1993891167, Iran
\end{center}
\vspace{2\baselineskip}
\begin{abstract}
\noindent The most general single species autonomous reaction-diffusion model on a Cayley
tree with nearest-neighbor interactions is introduced. The stationary solutions
of such models, as well as their dynamics, are discussed. To study dynamics of the system,
directionally-symmetric Green function for evolution equation of average number density
is obtained. In some limiting cases the Green function is studied. Some examples
are worked out in more detail.
\end{abstract}
\end{titlepage}
\newpage
\section{Introduction}
Most of the systems encountered in the nature are basically out of equilibrium.
Non-equilibrium systems have absorbed much interest recently. Different methods
have been used to study non-equilibrium systems, such as approximation methods,
simulation, mean-field techniques, and analytical methods. Most analytical studies
on reaction diffusion systems, however, belong to one dimensional systems.

The Cayley tree is a lattice without loops, where each site is connected
to $\xi$ neighbors, where $\xi$ is called the coordination number.
The Cayley tree with $\xi=2$ is in fact a one dimensional chain.
Cayley trees with $\xi \geq 3$ are far richer than the one dimensional chains.
The no-loop property of Cayley tree results in the solvability of
some models, for general coordination numbers. Reaction
diffusion models on the Cayley tree have been studied in, for example,
\cite{VNH,JKr,SNMP,KeSu,ACVP,SNMa,LAK,MANO,DBA,MMSKSEB}.
In \cite{VNH,JKr,SNMP} diffusion-limited aggregations,
and in \cite{SNMP} two-particle annihilation reactions
for immobile reactants have been studied. There are also
some exact results for deposition processes on the Bethe
lattice \cite{ACVP}.

In \cite{LAK} the most general reaction-diffusion model on
a Cayley tree with nearest-neighbor interactions was introduced,
which can be solved exactly through the empty-interval method.
The stationary solutions as well as the dynamics of such models
have been investigated there. In \cite{MANO}, It has been shown
that there exist two exactly solvable models. For the first model,
the probabilities of finding $m$ particles on the $l$-th shell of
the Cayley tree have been calculated. For the second model,
some other probabilities have been calculated.

Recently, a nonconsensus opinion model model on Bethe lattices
has been studied. It is argued there that the phase diagram
corresponding to such a model is different from that of regular
percolation \cite{DBA}. A class of cooperative sequential adsorption models
on a Cayley tree with constant and variable attachment rates
has been studied in \cite{MMSKSEB}. Possible applications
for ionic self-assembly of thin films and drug encapsulation
of nanoparticles have been also studied.

Most of the studies on autonomous reaction-diffusion systems
are on one-dimensional chains. Some of the results could be easily
generalized for multi-dimensional square lattices. A Cayley tree, on the other hand,
is an infinite-dimensional lattice, by which it is meant that
the number of sites the distances of which with a fixed site are less than
$a$ grows faster than any power of $a$.

Here a system is studied which consists of a Cayley tree with sites
being either occupied or empty. The evolution of the system is taken
to consist of nearest neighbor interaction, with the further
constraint that the system be autonomous, by which it is meant
that the evolution of densities (one-point functions) is closed.
The Green's function for the initial value problem of the densities
is calculated, and studied in more detail in some limiting cases.
Some special cases are also investigated more explicitly.
Especially, it is seen that the large-time behavior of the system
is different for the coordination number equal to $2$, and
the coordination number larger than $2$.

The scheme of the paper is as follows. In section 2, a system of
interacting particles on a Cayley tree is introduced.
The reactions are nearest neighbor interaction, and the rates are
such that the evolution equation of the average densities is closed.
The stationary solutions of such models are discussed.
In section 3, a Green's function approach is used to study
the dynamics of the system. The Green's function is calculated
and used to obtain, among other things, the large time behavior
of the system. Some examples are worked out in more detail.
Section 4 is devoted to the concluding remarks.
\section{The reactions}
The Cayley tree is a lattice without loops where
each site is connected to $\xi$ neighboring sites. Two sites are
called neighbors if they are connected through a link, and $\xi$  is called
the coordination number of the Cayley tree. 

\begin{picture}(100,130)(0,-10)
\includegraphics[scale=0.2]{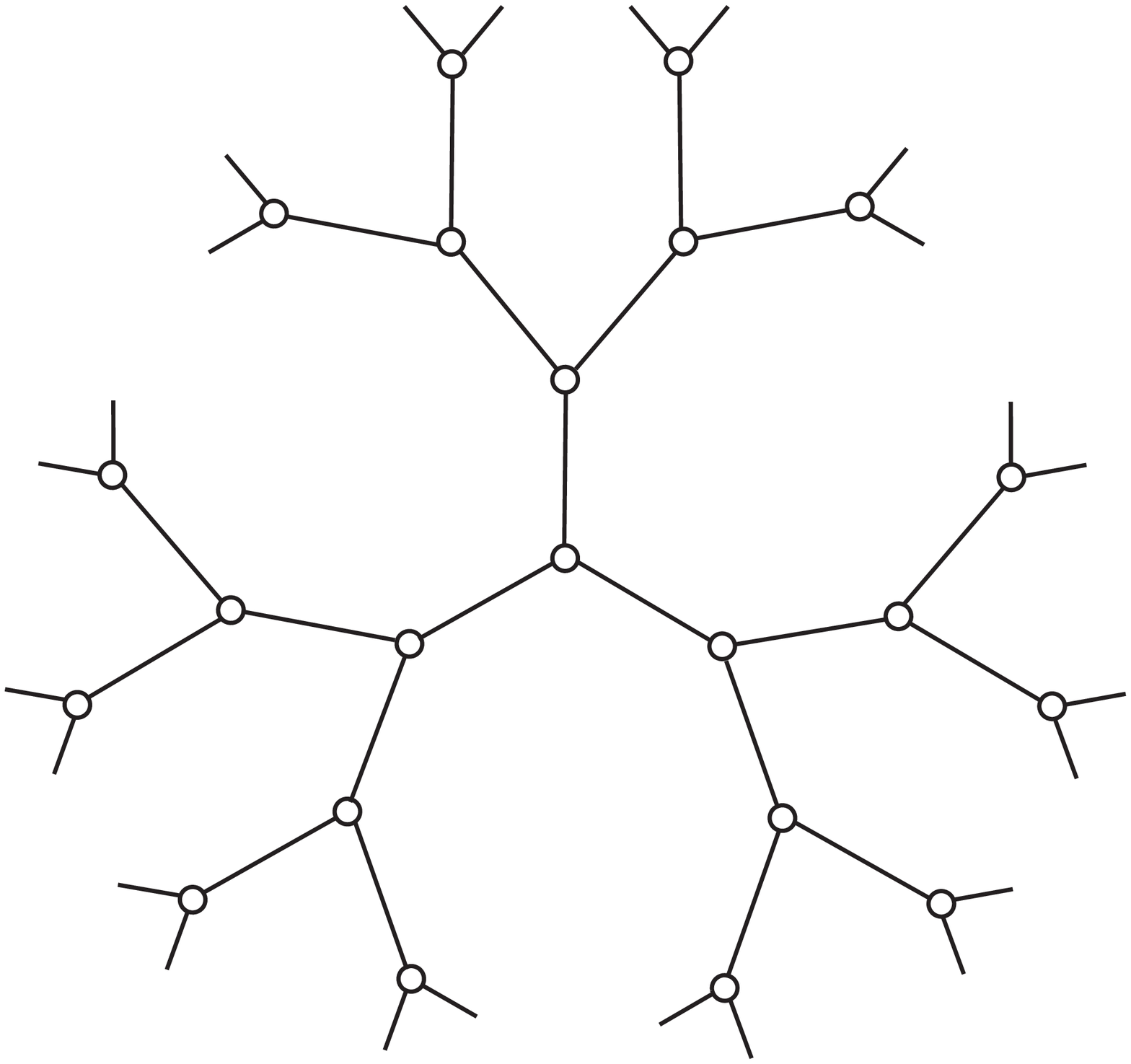}
\put(40,50){\small Part of a Cayley tree with $\xi=3$}
\end{picture}\\
Assume a system of interacting  particles on the Cayley tree. Each site is
either empty or occupied by one particle. The time evolution of
of each site depends on only that site and its nearest
neighbors, those which  directly related to it through a link.
Each site can be either occupied, denoted by $\bullet$, or vacant, denoted by $\circ$.
The number operator corresponding to the site $i$ is denoted by
$n_i$, which is $0$ ($1$), if the site $i$ is vacant (occupied).
The evolution of such a system is governed by a Hamiltonian $H$.
Two neighboring sites with the joint state $(\nu\,\mu)$  can evolve
to another joint state $(\lambda\,\kappa) \ne (\nu\,\mu) $
with the rate $H^{\lambda\,\kappa}{}_{\nu\,\mu}$.

A system is called autonomous when the evolution of the average
density, $\langle n_i \rangle$, is closed, by which it is meant
that the evolution of $\langle n_i \rangle$ is expressible
in terms of only $\langle n_j \rangle$'s. The criterion for autonomy
has been obtained in \cite{sch}. The argument is that corresponding
to the site $i$, the time derivative of $\langle n_i\rangle$ is a summation
of terms corresponding to the links which contain $i$:
\begin{equation}
\frac{\rd}{\rd t}\langle n_i\rangle=\sum_{j\in\mathsf{N}_1(i)}
\left(\frac{\rd}{\rd t}\langle n_i\rangle\right)_j,
\end{equation}
where $\mathsf{N}_a(i)$ is the set of sites the distance of them from
the site $i$ is $a$, and
\begin{align}
\left(\frac{\rd}{\rd t}\langle n_i\rangle\right)_j&=
+(H^{1\,0}{}_{0\,0}+H^{1\,1}{}_{0\,0})\,\langle(1-n_i)\,(1-n_j)\rangle\nonumber\\
&\quad+(H^{1\,0}{}_{0\,1}+H^{1\,1}{}_{0\,1})\,\langle(1-n_i)\,n_j\rangle\nonumber\\
&\quad-(H^{0\,0}{}_{1\,0}+H^{0\,1}{}_{1\,0})\,\langle n_i\,(1-n_j)\rangle\nonumber\\
&\quad-(H^{0\,0}{}_{1\,1}+H^{0\,1}{}_{1\,1})\,\langle n_i\,n_j\rangle.
\end{align}
The condition for autonomy is that the coefficient of $\langle n_i\,n_j\rangle$
in the right-hand side vanishes. Assuming that the interactions in
each link are symmetric:
\begin{equation}
H^{\lambda\,\kappa}{}_{\nu\,\mu}=H^{\kappa\,\lambda}{}_{\mu\,\nu},
\end{equation}
one arrives at the following criterion for the autonomy:
\begin{equation}\label{autonomy}
H^{0\,0}{}_{1\,1}+H^{1\,1}{}_{0\,1}+H^{0\,1}{}_{1\,1}=H^{1\,1}{}_{0\,0}
+H^{0\,0}{}_{1\,0}+H^{1\,0}{}_{0\,0}.
\end{equation}

From now on, it is assumed that the interactions in each link are symmetric,
and the autonomy criterion holds. Defining
\begin{equation}
\bm{\rho}_i:=\langle n_i\rangle,
\end{equation}
the evolution for $\bm{\rho}$ would be
\begin{equation}\label{evo}
\dot{\bm{\rho}}_i=\alpha\,\xi+\sum_{j\in\mathsf{N}_1(i)}(\beta\,\bm{\rho}_j-\gamma\,\bm{\rho}_i),
\end{equation}
where $\xi$ is the number of neighbors to each site, and
\begin{align}
\alpha&:=H^{1\,1}{}_{0\,0}+H^{1\,0}{}_{0\,0},\nonumber\\
\beta&:=H^{0\,1}{}_{1\,0}+H^{1\,1}{}_{1\,0}-(H^{1\,1}{}_{0\,0}+H^{1\,0}{}_{0\,0}),\nonumber\\
\gamma&:=H^{0\,1}{}_{1\,0}+H^{0\,0}{}_{0\,1}+(H^{1\,1}{}_{0\,0}+H^{1\,0}{}_{0\,0}).
\end{align}
The requirement that the rates be nonnegative, and the criterion (\ref{autonomy})
for the autonomy, lead to the following constraints
\begin{align}
\alpha&\geq0,\nonumber\\
\beta&\geq-\alpha,\nonumber\\
\gamma&\geq\alpha,\nonumber\\
\gamma&\geq\alpha+\beta.
\end{align}

Equation (\ref{evo}) which describes the evolution of the density, is clearly 
translational invariant. The initial condition, however, is not necessarily 
so. So in general the density does depend on the site. 
At large times, system approaches to its stationary state, 
which is  a state with uniform density.
The stationary solution to the above evolution is $\rho^\mathrm{st}$ is
\begin{equation}
\bm{\rho}^\mathrm{st}=\frac{\alpha}{\gamma-\beta}.
\end{equation}
Defining the dynamic solution $\bm{\rho}^\mathrm{dy}$ through
\begin{equation}
\bm{\rho}=:\bm{\rho}^\mathrm{st}+\bm{\rho}^\mathrm{dy},
\end{equation}
one arrives at
\begin{align}\label{rdy}
\dot{\bm{\rho}}_i^\mathrm{dy}&=
\sum_{j\in\mathsf{N}_1(i)}(\beta\,\bm{\rho}_j^\mathrm{dy}-\gamma\,\bm{\rho}_i^\mathrm{dy}),\\
&=:\bm{h}_i{}^j\,\bm{\rho}_j^\mathrm{dy}.
\end{align}

If the initial conditions are directionally-symmetric, by which it
is meant that $\bm{\rho}_i$ depends on only the distance of $i$ from
some center, then the density remains directionally-symmetric.
In such a situation, one defines another density $\rho$ through
\begin{equation}
\rho_a:=\bm{\rho}_i,\qquad i\in\mathsf{N}_a(0).
\end{equation}
The evolution of this would be
\begin{align}
\dot\rho_0&=\alpha\,\xi-\gamma\,\xi\,\rho_0+\beta\,\xi\,\rho_1,\nonumber\\
\dot\rho_a&=\alpha\,\xi+\beta\,\rho_{a-1}-\gamma\,\xi\,\rho_0
+\beta\,(\xi-1)\,\rho_{a+1},\qquad a>0.
\end{align}
The case $(a=0)$ can be included in the general case, if one defines
\begin{equation}\label{bo}
\rho_{-1}:=\rho_1.
\end{equation}
Again, for the difference of $\varrho$ and its stationary value, one has
\begin{align}\label{rrdy}
\dot\rho_0^\mathrm{dy}&=-\gamma\,\xi\,\rho_0^\mathrm{dy}
+\beta\,\xi\,\rho_1^\mathrm{dy},\nonumber\\
\dot\rho_a^\mathrm{dy}&=\beta\,\rho_{a-1}^\mathrm{dy}
-\gamma\,\xi\,\rho_a^\mathrm{dy}+\beta\,(\xi-1)\,\rho_{a+1}^\mathrm{dy},\qquad a>0,
\end{align}
which can be written as
\begin{equation}\label{rsy}
\dot\rho_a^\mathrm{dy}=h_a{}^b\,\rho_b^\mathrm{dy},
\end{equation}
where
\begin{align}
h_0{}^b&=-\gamma\,\xi\,\delta_0^b+\beta\,\xi\,\delta_1^b,\nonumber\\
h_a{}^b&=\beta\,\delta_a^{b+1}-\gamma\,\xi\,\delta_a^b+\beta\,(\xi-1)\,\delta_a^{b-1},
\qquad a>0.
\end{align}
\section{The initial value problem}
To solve the initial value problem for the dynamic solution, one way is to
use the Green's function method. Denoting the Green's function by
$\bm{G}$, one would have
\begin{align}
\bm{\rho}_i^\mathrm{dy}(t)&=\sum_j\bm{G}_i{}^j(t)\,\bm{\rho}_j^\mathrm{dy}(0),\\
\dot{\bm{G}}_i{}^j(t)&=\bm{h}_i{}^l\,\bm{G}_l{}^j(t),\\
\bm{G}_i{}^j(0)&=\delta_i^j.
\end{align}
The initial value configuration for the Green's function
$\bm{G}$ is directionally-symmetric (with respect to the point $j$). So
the evolution of $\bm{G}$ is governed by an equation similar to
(\ref{rsy}). To be more specific,
\begin{equation}
\bm{G}_i{}^j=G_{\mathrm{d}(i,j)}{}^0,
\end{equation}
where $\mathrm{d}(i,j)$ is the distance of $i$ from $j$, and $G$ satisfies
\begin{align}
\dot G_a{}^b(t)&=h_a{}^c\,G_c{}^b(t),\\
G_a{}^b(0)&=\delta_a^b.
\end{align}
The solution to the initial value problem for $\rho^\mathrm{dy}$ would then be
\begin{equation}
\bm{\rho}^\mathrm{dy}_i(t)=\sum_j G_{\mathrm{d}(i,j)}{}^0(t)\,\bm{\rho}^\mathrm{dy}_j(0).
\end{equation}
Of course $G$ can also be used to solve the directionally-symmetric initial value problem:
\begin{equation}\label{dir}
\rho^\mathrm{dy}_a(t)=G_a{}^b(t)\,\rho_b(0).
\end{equation}
Comparing these, one arrives at
\begin{equation}
G_a{}^b=\sum_{j\in\mathsf{N}_b(0)}\bm{G}_i{}^j,\qquad i\in\mathsf{N}_a(0).
\end{equation}

To find the Green's function $G$, one could use the usual
method of finding the eigenvectors of $h$:
\begin{equation}
h\,\psi_E=E\,\psi_E,
\end{equation}
where $\psi_E$ is the eigenvector of $h$ corresponding to the eigenvalue $E$.
An ansatz for $\psi_E$ is
\begin{equation}
\psi_{E\,a}=\sum_\nu C_\nu\,(z_\nu)^a,
\end{equation}
where $z_\nu$'s are the roots of the function $(f-E)$ with
\begin{equation}
f(z):=\frac{\beta}{z}+\beta\,(\xi-1)\,z-\gamma\,\xi.
\end{equation}
The function $(f-E)$, has obviously two roots, and
\begin{equation}\label{pro}
z_1\,z_2=\frac{1}{\xi-1}.
\end{equation}
The analog of (\ref{bo}) for $\psi_E$, then results in
\begin{equation}\label{fir}
C_1\,\left(z_1-\frac{1}{z_1}\right)+C_2\,\left(z_2-\frac{1}{z_2}\right)=0.
\end{equation}
There is another boundary condition, which is $\varrho_a$ should not blow up
at $a\to\infty$. This can be realized by introducing some $N$ and demanding
$\varrho_N$ be zero, and then sending $N$ to infinity. So the other boundary
condition would be
\begin{equation}\label{la}
C_1\,(z_1)^N+C_2\,(z_2)^N=0.
\end{equation}
If $|z_1|>|z_2|$, the above boundary condition would result in
the vanishing of $C_2$ at $N\to\infty$;
and (\ref{pro}) would mean that $|z_2|<1$, so that (\ref{fir})
would result in $C_2$ being zero too. So in order that $\psi_E$ be non vanishing,
\begin{equation}
|z_2|=|z_1|,
\end{equation}
which, defining
\begin{equation}
\eta:=\frac{1}{2}\,\ln(\xi-1),
\end{equation}
results in
\begin{align}
z_{1,2}(\theta)&=\frac{1}{\sqrt{\xi-1}}\,\exp(\pm\ri\,\theta),\nonumber\\
&=\exp(-\eta\pm\ri\,\theta),\\
E(\theta)&=2\,\beta\,\exp(\eta)\,\cos\theta-\gamma\,\xi,\nonumber\\
&=2\,\beta\,\exp(\eta)\,\cos\theta-2\,\gamma\,\exp(\eta)\,\cosh\eta,\\
\psi_{\theta\,a}&=\sinh(\eta+\ri\theta)\,\exp[a\,(-\eta+\ri\,\theta)]
-\sinh(\eta-\ri\theta)\,\exp[a\,(-\eta-\ri\,\theta)],
\end{align}
where (\ref{fir}) has been used and a change of notation has been made from
$\psi_E$ to $\psi_\theta$. It is seen that $\eta$ is nonnegative, and it is
zero only for $\xi=2$, which corresponds to a one-dimensional chain
(instead of the Cayley tree).

To solve the initial value problem for $\varrho^\mathrm{dy}$, one
expands the $\varrho^\mathrm{dy}(0)$ in terms of the eigenvectors
of $h$. To do so, it is helpful to find the right eigenvectors of $h$.
Denoting these by $\phi$, one has
\begin{align}
(\phi\,h)^0&=-\gamma\,\xi\,\phi^0+\beta\,\phi^1,\nonumber\\
(\phi\,h)^1&=\beta\,\xi\,\phi^0-\gamma\,\xi\,\phi^1+\beta\,\phi^2,\nonumber\\
(\phi\,h)^b&=\beta\,(\xi-1)\,\phi^{b-1}-\gamma\,\xi\,\phi^b+\beta\,\phi^{b+1},\qquad b>1.
\end{align}
So $\phi_\theta$, the right eigenvector of $h$ corresponding to
the eigenvalue $E(\theta)$, would satisfy
\begin{align}
[E(\theta)]\,\phi_\theta^0&=-\gamma\,\xi\,\phi_\theta^0+\beta\,\phi_\theta^1,\nonumber\\
[E(\theta)]\,\phi_\theta^1&=\beta\,\xi\,\phi_\theta^0-\gamma\,\xi\,\phi_\theta^1
+\beta\,\phi_\theta^2,\nonumber\\
[E(\theta)]\,\phi_\theta^b&=\beta\,(\xi-1)\,\phi_\theta^{b-1}-\gamma\,\xi\,\phi_\theta^b
+\beta\,\phi_\theta^{b+1},\qquad b>1.
\end{align}
The last equation suggests
\begin{equation}
\phi_\theta^b=D_1\,(z_1)^{-b}+D_2\,(z_2)^{-b},\qquad b>0.
\end{equation}
Putting this in the remaining equations, one arrives at
\begin{align}
\phi_\theta^0&=\frac{\xi-1}{\xi}\,(D_1+D_2),\nonumber\\
\xi\,\left(\frac{D_1}{z_1}+\frac{D_2}{z_2}\right)&=(\xi-1)\,(D_1+D_2)
\,\left(\frac{1}{z_1}+\frac{1}{z_2}\right).
\end{align}
So,
\begin{equation}
\phi_\theta^b=A\,\left(1-\frac{1}{\xi}\,\delta_0^b\right)\,\{
\sinh(\eta-\ri\theta)\,\exp[b\,(\eta-\ri\,\theta)]
-\sinh(\eta+\ri\theta)\,\exp[b\,(\eta+\ri\,\theta)]\},
\end{equation}
where $A$ is some normalization constant. Using
\begin{equation}
\sum_{a=0}^\infty\exp(\ri\,a\,\chi)=\mathrm{pf}\left[\frac{1}{1-\exp(\ri\,\chi)}\right]
+\pi\,\delta(\chi),
\end{equation}
one obtains
\begin{equation}
\sum_{a=0}^\infty\phi_\theta^a\,\psi_{\theta'\,a}=2\,\pi\,A
\,\sinh(\eta+\ri\,\theta)\,\sinh(\eta-\ri\,\theta)\,[\delta(\theta-\theta')
-\delta(\theta+\theta')].
\end{equation}
So the left eigenvectors are normalized through
\begin{equation}
A=\frac{1}{2\,\pi\,\sinh(\eta+\ri\,\theta)\,\sinh(\eta-\ri\,\theta)},
\end{equation}
resulting in
\begin{equation}
\phi_\theta^b=\frac{1}{2\,\pi}\,\left[1-\frac{\exp(-\eta)}{2\,\cosh\eta}\,\delta_0^b\right]\,
\left\{\frac{\exp[b\,(\eta-\ri\,\theta)]}{\sinh(\eta+\ri\theta)}-
\frac{\exp[b\,(\eta+\ri\,\theta)]}{\sinh(\eta-\ri\theta)}\right\},
\end{equation}
with
\begin{equation}
\sum_{a=0}^\infty\phi_\theta^a\,\psi_{\theta'\,a}=\delta(\theta-\theta')-\delta(\theta+\theta').
\end{equation}
Using these, the Green's function for the evolution (\ref{rrdy}) is seen to be
\begin{align}\label{gre}
G_a{}^b(t)&=\int_0^\pi\rd\theta\,\psi_{\theta\,a}\,\phi_\theta^b\,\exp[t\,E(\theta)],\nonumber\\
&=\frac{1}{2\,\pi}\,\left[1-\frac{\exp(-\eta)}{2\,\cosh\eta}\,\delta_0^b\right]\,
\exp\{(b-a)\,\eta-[2\,\gamma\,\cosh\eta\,\exp(\eta)]\,t\}\nonumber\\
&\quad\times\int_{-\pi}^\pi\rd\theta\,\left\{\exp[\ri\,(a-b)\,\theta]+
\frac{\sinh(\ri\,\theta+\eta)}{\sinh(\ri\,\theta-\eta)}\,\exp[\ri\,(a+b)\,\theta]\right\}\nonumber\\
&\quad\times\exp\{[2\,\beta\,\cos\theta\,\exp(\eta)]\,t\}.
\end{align}

Among other things, one can investigate some special or limiting cases.
\subsection{Large time limit} It is seen that at large times the integral is dominated by
the integrand around $\theta=0$. Defining
\begin{equation}
B(\theta):=\exp[\ri\,(a-b)\,\theta]+
\frac{\sinh(\ri\,\theta+\eta)}{\sinh(\ri\,\theta-\eta)}\,\exp[\ri\,(a+b)\,\theta],
\end{equation}
it is seen that the behavior of $B$ for small $\theta$ is different for
$\eta$ being zero or positive. In fact only the even part of $B$ contributes
to the integral, and that part behaves (for small $\theta$) as
\begin{equation}
\frac{B(\theta)+B(-\theta)}{2}=\begin{cases}
2+\cdots&\eta=0,\\
2\,(a+\coth\eta)\,(b+\coth\eta)\,\theta^2+\cdots,&\eta>0,
\end{cases}.
\end{equation}
So at large times,
\begin{alignat}{2}
G_a{}^b(t)&=\frac{1}{\sqrt{\pi\,\beta\,t}}\,\left(1-\frac{1}{2}\,\delta_0^b\right)\,
\exp[2\,(\beta-\gamma)\,t],&\qquad&\eta=0.\\
G_a{}^b(t)&=\frac{(a+\coth\eta)\,(b+\coth\eta)}{2\,\sqrt{\pi\,[\beta\,\exp(\eta)\,t]^3}}\,
\left[1-\frac{\exp(-\eta)}{2\,\cosh\eta}\,\delta_0^b\right]\nonumber\\
&\quad\times\exp\{(b-a)\,\eta+[2\,(\beta-\gamma\,\cosh\eta)\,\exp(\eta)]\,t\},&\qquad&\eta>0.
\end{alignat}
It is seen that the large time behavior of the system for
$\eta=0$ (equivalent to $\xi=2$, when the Cayley tree is
a one dimensional chain) is different from that of $\eta>0$
(corresponding to $\xi>2$). Even the large time behavior for $\eta\to 0$
is different from the large time behavior for $\eta=0$.
The reason is that in approximating $B(\theta)$ for small values
of $\theta$, the result is different for $\eta\to 0$
and $\eta=0$, due to the fraction in the second term. For $\eta=0$, that fraction
is equal to $1$, no matter how small $\theta$ is. But if $\eta>0$, no matter how small
it is, there are values for $\theta$ which are much smaller than $\eta$, which make
the fraction equal to $(-1)$. One can, however, find a cross-over time when
such a shift in behavior occurs. This comes from the fact that for large values
of $t$, the relevant values of $\theta$ are those which are less than a ceratin
value $\theta_0$:
\begin{equation}
\theta_0\sim(\beta\,t)^{-1/2}.
\end{equation}
If $\theta_0$ is much smaller than $\eta$, then the behavior corresponds
to the case of $\eta>0$. So the cross-over time $t_\mathrm{c}$ satisfies
\begin{equation}
t_\mathrm{c}\sim(\beta\,\eta^2)^{-1}.
\end{equation}
If $t$ is large but still much smaller than $t_\mathrm{c}$, then the behavior is
similar to the case of $\eta=0$.
\subsection{One dimensional chain}
The one dimensional chain is a Cayley tree with $\xi=2$
(equivalently $\eta=0$). In that case, the integration in the expression for
the Green's function is readily performed and one obtains
\begin{equation}
G_a{}^b(t)=\left(1-\frac{1}{2}\,\delta_0^b\right)\,\exp(-2\,\gamma\,t)\,
[\mathrm{I}_{a-b}(2\,\beta\,t)+\mathrm{I}_{a+b}(2\,\beta\,t)],\qquad\xi=2,
\end{equation}
where $\mathrm{I}_c$ is the modified Bessel function of the first kind of order $c$.\\
\subsection{Highly connected Cayley tree}
The Green's function is simplified in the opposite limit (large $\xi$)
as well. For large values of $\xi$ (or $\eta$), the integral in
the expression of the Green's function is again dominated by
the value of the integrand for small $\theta$. Hence a techinque similar
to what was used for the large time behavior can be used here. One arrives at
\begin{align}
G_a{}^b(t)&=\frac{(a+1)\,(b+1)}{2\,\sqrt{\pi\,[\beta\,\exp(\eta)\,t]^3}}\,
\exp\{(b-a)\,\eta+[2\,\beta\,\exp(\eta)-\gamma\,\exp(2\,\eta)]\,t\},\nonumber\\
&\qquad\eta\gg 1\mbox{ (equivalent to $\xi\gg 2$)}.
\end{align}
\subsection{directionally-symmetric initial conditions}
If the initial density is directionally-symmetric, then (\ref{dir})
can be used to obtain the density at the time $t$. As an example, consider
\begin{equation}
\rho^\mathrm{dy}_a(0)=\begin{cases}\varrho,& a\leq r\\
0,& a>r
\end{cases},
\end{equation}
where $\varrho$ is a constant. One arrives at
\begin{equation}
\rho^\mathrm{dy}_a(t)=\varrho\,\sum_{b=0}^r G_a{}^b(t).
\end{equation}
So,
\begin{align}
\rho^\mathrm{dy}_a(t)&=\frac{\varrho}{2\,\pi}\,\exp\{-a\,\eta-[2\,\gamma\,\cosh\eta\,\exp(\eta)]\,t\}
\nonumber\\
&\quad\times\Bigg(-\frac{\exp(-\eta)}{2\,\cosh\eta}\,\int_{-\pi}^\pi\rd\theta\,\exp(\ri\,a\,\theta)\,
\bigg[1+\frac{\sinh(\ri\,\theta+\eta)}{\sinh(\ri\,\theta-\eta)}\bigg]\nonumber\\
&\quad+\int_{-\pi}^\pi\rd\theta\,\bigg\{
\frac{\exp[(r+1)\,(\eta-\ri\,\theta)]-1}{\exp(\eta-\ri\,\theta)-1}\nonumber\\
&\quad+
\frac{\exp[(r+1)\,(\eta+\ri\,\theta)]-1}{\exp(\eta+\ri\,\theta)-1}\,
\frac{\sinh(\ri\,\theta+\eta)}{\sinh(\ri\,\theta-\eta)}\bigg\}\,\exp(\ri\,a\,\theta)\Bigg)\nonumber\\
&\quad\times\exp\{[2\,\beta\,\cos\theta\,\exp(\eta)]\,t\},
\end{align}
where (\ref{gre}) has been used.
\section{Concluding remarks}
A system of particles was studied which move and react
on a cayley tree, so that the interactions are nearest-neighbor, and
autonomous. The Green's function for the initial value problem of
the densities was calculated, and its behavior for large times, as well as
small and large coordination numbers was studied. It was seen that
the large-time behavior of the system is different for t
he coordination number equal to $2$, and the coordination number
larger than $2$.
\\[\baselineskip]
\textbf{Acknowledgement}: This work was supported by
the research council of the Alzahra University.

\end{document}